# Nanowire Volatile RAM as an Alternative to SRAM

Mostafizur Rahman, Santosh Khasanvis, and Csaba Andras Moritz
Dept. of Electrical and Computer Engineering
University of Massachusetts Amherst, Amherst, USA
{rahman, andras}@ecs.umass.edu

***Abstract*— Maintaining benefits of CMOS technology scaling is becoming challenging due to increased manufacturing complexities and unwanted passive power dissipations. This is particularly challenging in SRAM, where manufacturing precision and leakage power control are critical issues. To alleviate some of these challenges a novel non-volatile memory alternative to SRAM was proposed called nanowire volatile RAM (NWRAM) [1]. Due to NWRAMs regular grid based layout and innovative circuit style, manufacturing complexity is reduced and at the same time considerable benefits are attained in terms of performance and leakage power reduction. In this paper, we elaborate more on NWRAM circuit aspects and manufacturability, and quantify benefits at 16nm technology node through simulation against state-of-the-art 6T-SRAM and gridded 8T-SRAM designs. Our results show the 10T-NWRAM to be 2x faster and 35x better in terms of leakage when compared to high performance gridded 8T-SRAM design.**

## I. Introduction

The continuous push for denser, faster and more power efficient computing is driving CMOS scaling to its limit. Numerous new challenges are emerging related to power consumption, circuit noise, manufacturability and cost. These challenges are especially critical for CMOS SRAM circuits, where both PMOS and NMOS transistors need to be precisely sized and doped for memory operation and for sufficient noise margin. Due to the complex and compact layout of SRAM circuits, it is becoming difficult to maintain such precision at nanoscale. Moreover, controlling passive power in SRAM circuits is becoming a big concern; this is mainly because of the static SRAM circuit style and leakage current increase in nanoscale transistors.

To overcome these issues in SRAM, we recently proposed a novel nanowire based volatile RAM (NWRAM) [1]. It alleviates manufacturing requirements through a very regular grid based layout, single-type and uniformly-sized transistors, and fewer metal layers. In contrast to SRAM's inverter based latching approach that requires careful transistor sizing for write-ability and read stability, NWRAM employs uniformly sized transistors in dynamic circuits with non-overlapping clocking control scheme and synchronous data input to achieve memory functionality. In addition, the read logic in NWRAM is separated from bit storage, which avoids read-stability concerns present in SRAM. Furthermore, NWRAMs dynamic circuit approach with multiple nanowire transistors between storage and sink allows significant leakage control.

In this paper, we provide an in-depth discussion on NWRAMs circuit and manufacturing aspects; we show detailed methodology and benchmarking of a 10 transistor NWRAM (10T-NWRAM) against state-of-the-art 6 transistor SRAM (6T-SRAM) and 8 transistor based gridded SRAM (8T-SRAM) designs, in 16nm technology node. Our results show that 10T-NWRAM is 2x faster compared to high performance 8T-SRAM, and consumes 35x less leakage power with respect to high performance 8T-SRAM while maintaining comparable density benefits, with lesser manufacturing requirements. This paper is organized as follows: Section II presents underlying nanowire based fabric for 10T-NWRAM, Section III discusses 10T-NWRAM circuit and layout details, Section IV shows benchmarking methodology and results, and Section V concludes the paper.

## II. Underlying Fabric: N$^3$ASIC

N$^3$ASIC [2] is a physical fabric where nanoscale devices and interconnects are integrated in a novel manner, which enables efficient logic and memory implementations. In this fabric, manufacturing complexities are reduced through the use of regular grid-based layout and uniform devices with no arbitrary device-sizing. This allow a combination of unconventional manufacturing (e.g., nanoimprint) and conventional lithography approaches (following lithographic design rules) to be leveraged for realizing high density nanofabrics with relaxed overlay precision requirements [1].

As shown in Fig. 1, N$^3$ASIC building blocks are: arrays of patterned semiconductor nanowires, cross-nanowire FETs (xnwFETs), orthogonal metal gate inputs, control and power rails, standard vias and the 3-D metal stack. All logic and memory functionalities are achieved in semiconductor nanowires. Depending on the functionalities, xnwFETs are placed on certain cross-points between input metal gate and bottom nanowire, vias carry output of each logic stage, and the input/output signals are routed through 3-D metal stack.

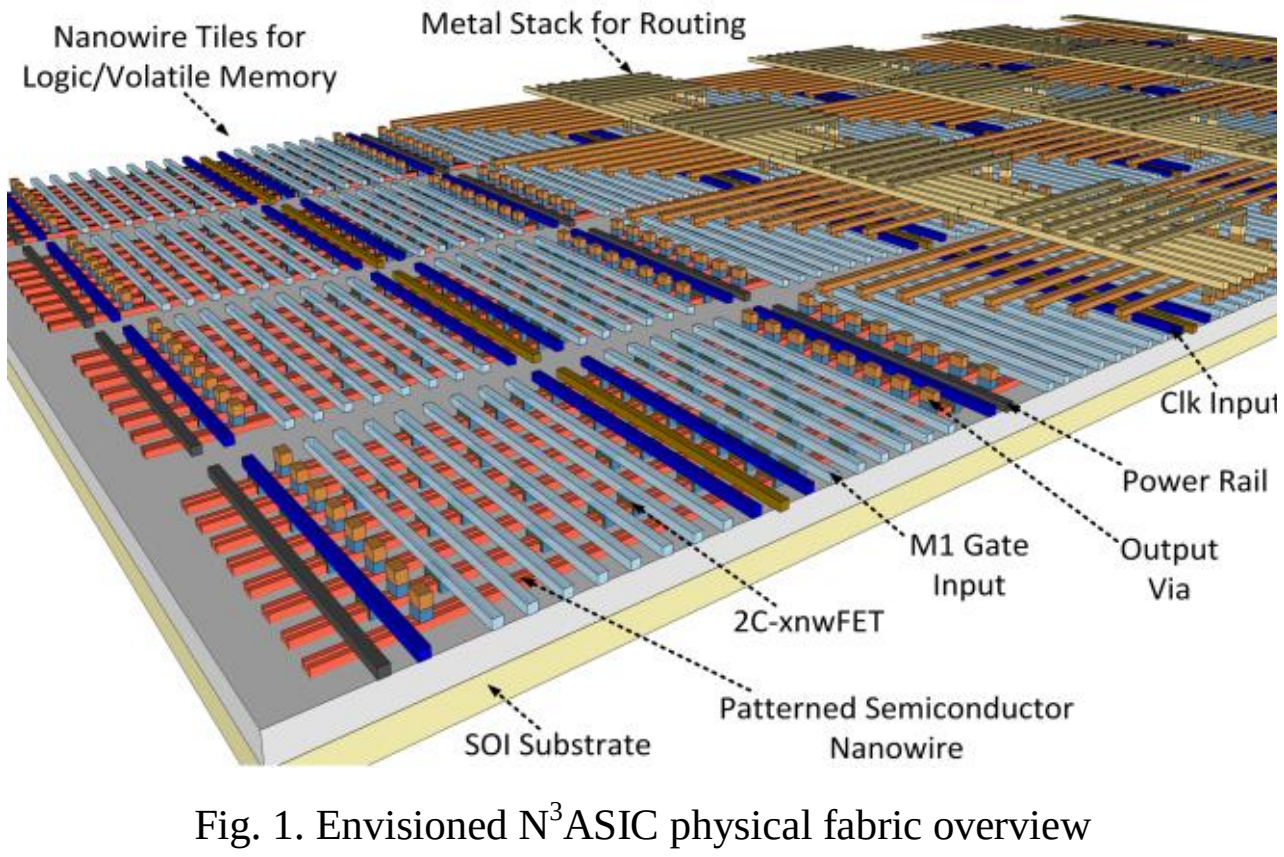

Fig. 1. Envisioned N$^3$ASIC physical fabric overview

In $N^3$ASIC a dynamic circuit style [2] that is amenable to implementation in regular nanowire arrays is used. All the devices are of same type and uniform size; active devices are N-type dual channel cross-nanowire FETs (2C-xnwFETs). Our 10T-NWRAM implementation uses this $N^3$ASIC framework and follows a similar circuit style.

## III. 10T-NWRAM

The core of 10T-NWRAM circuit consists of two cross-coupled dynamic NAND gates that store true and complementary data values on their outputs. In order to read-out the stored value, a separate read path is used. A set of select clock inputs ($W_0pre_0$, $W_0pre_1$, $W_0eva_0$, $W_0eva_1$), synchronous data input ($bit_0$) and a read signal ($read_0$) is used for memory operations (write, read and restore).

10T-NWRAM circuit schematic is shown in Fig. 2. The non-overlapping clock control signals ($W_0pre_0$, $W_0pre_1$, $W_0eva_0$, $W_0eva_1$) serve as word select, and are used to write data input ($bit_0$) in the form of true (*out*) and complementary values (*nout*) at first row ($W_o$) and first column ($bit_0$) position of the memory array. The read signal ($read_0$) is used to read memory output ($bit_0$, $bit_{1,...,}$ $bit_n$) from first row. Fig. 3 shows 10T-NWRAM array organization.

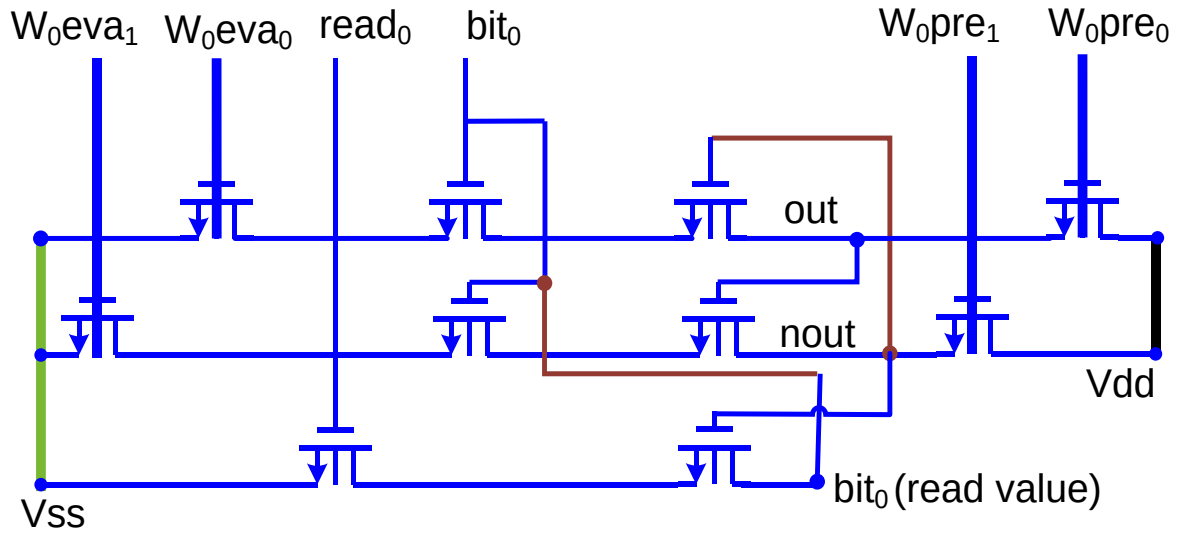

Fig. 2. 10-T NWRAM circuit schematic

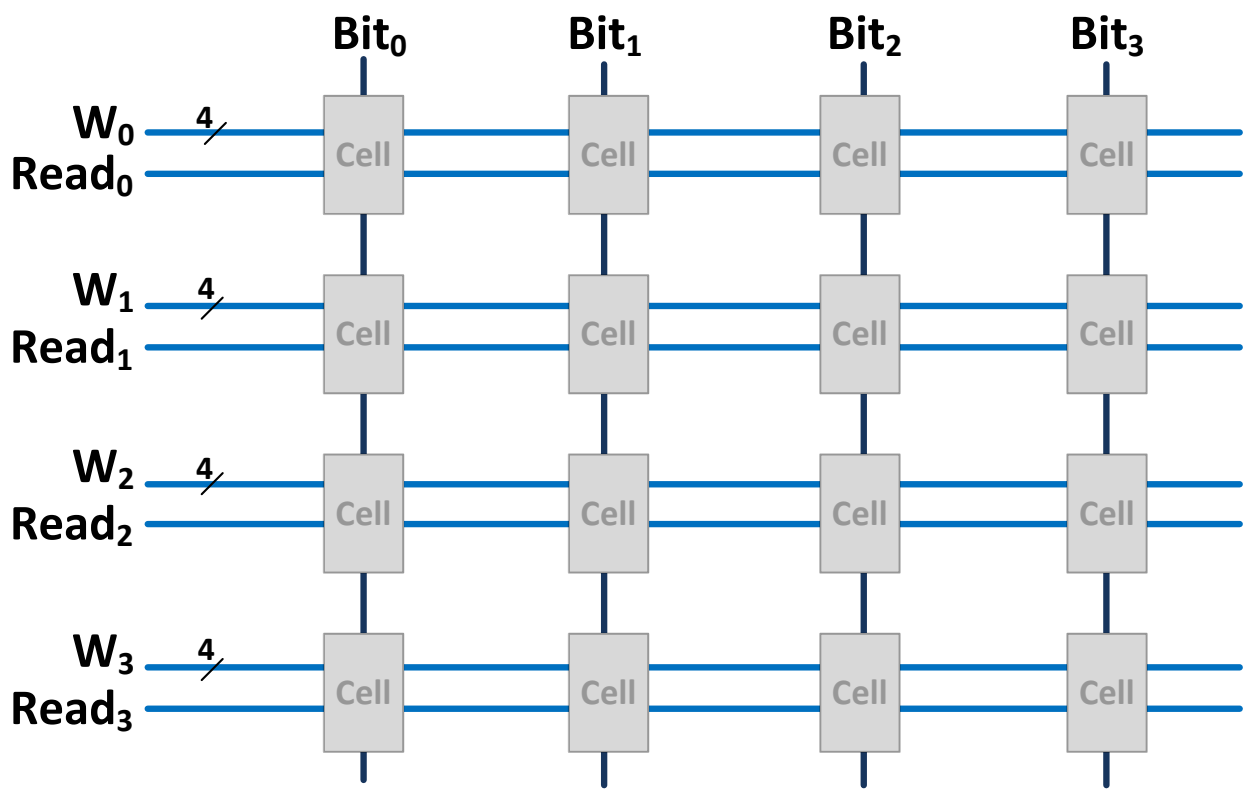

Fig. 3. 4 x 4 NWRAM array organization

In the following we discuss the basic memory operations in a single 10T-NWRAM cell.

Write operation in 10T-NWRAM is performed by synchronizing the $bit_0$ signal with either ($W_0pre_0$, $W_0eva_0$) or ($W_0pre_1$, $W_0eva_1$) clock pair signals. For example, to write '1' (i.e., *out* = 1, *nout* = 0) in the memory cell, $bit_0$ is kept low during precharge ($W_0pre_0$) and evaluate ($W_0eva_0$) phase of the corresponding NAND gate; as a result *out* retains its precharged value '1'. During subsequent precharge ($W_0pre_1$) and evaluate ($W_0eva_1$) phases of second NAND gate $bit_0$ is kept high, as a result *nout* becomes '0' storing the complement of *out*. In order to write '0' to *out*, the opposite sequence is followed by pulling up *nout* to '1' first. HSPICE simulation results are shown in fig. 4., which validates the concept. Once the *out* and *nout* values are set, they are retained in subsequent clock cycles due to the self-restoring nature of this cross-coupled circuit.

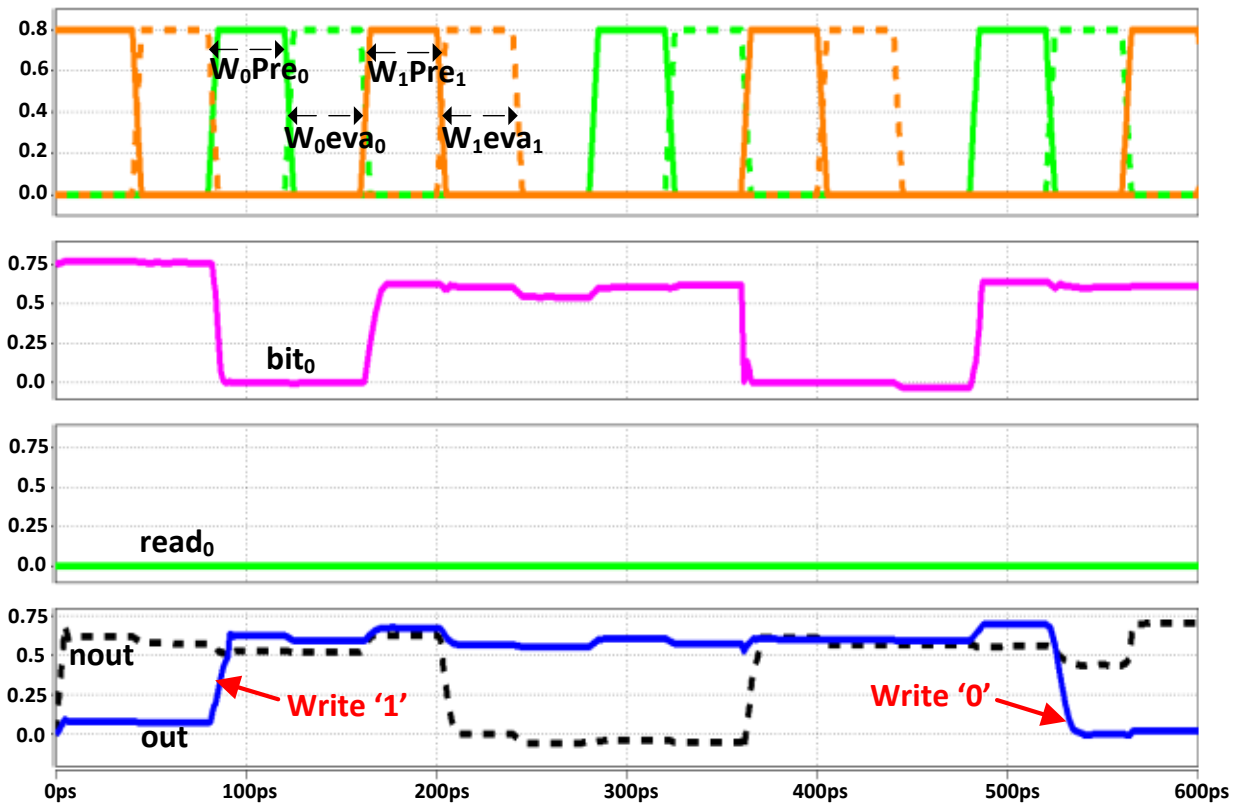

Fig. 4. Simulated waveforms of 10T-NWRAM operations

Read operation is achieved by gating the *nout* signal in a 2-input dynamic NAND gate. The signal $bit_0$ is used to carry the read output, since it is shared across multiple cells in a column of NWRAM array (Fig. 3). In order to perform read operation, $bit_0$ is initially precharged to '1', and then the $read_0$ signal is turned *ON*; depending on the value stored in *nout*, $bit_0$ is either pulled to $V_{ss}$ or kept high, performing the readout of a stored bit. The read operation is illustrated in fig. 5 through simulated waveforms. Fig. 5 shows both read '0' and read '1' operations. When the stored bit is '0' (i.e., *out* = 0, *nout* = 1) and $read_0$ signal is high, $bit_0$ signal is pulled to $V_{ss}$. On the other hand, when the stored bit is '1' (i.e. *out* = 1, *nout* = 0) and $read_0$ signal is high, $bit_0$ remains high at $V_{dd}$.

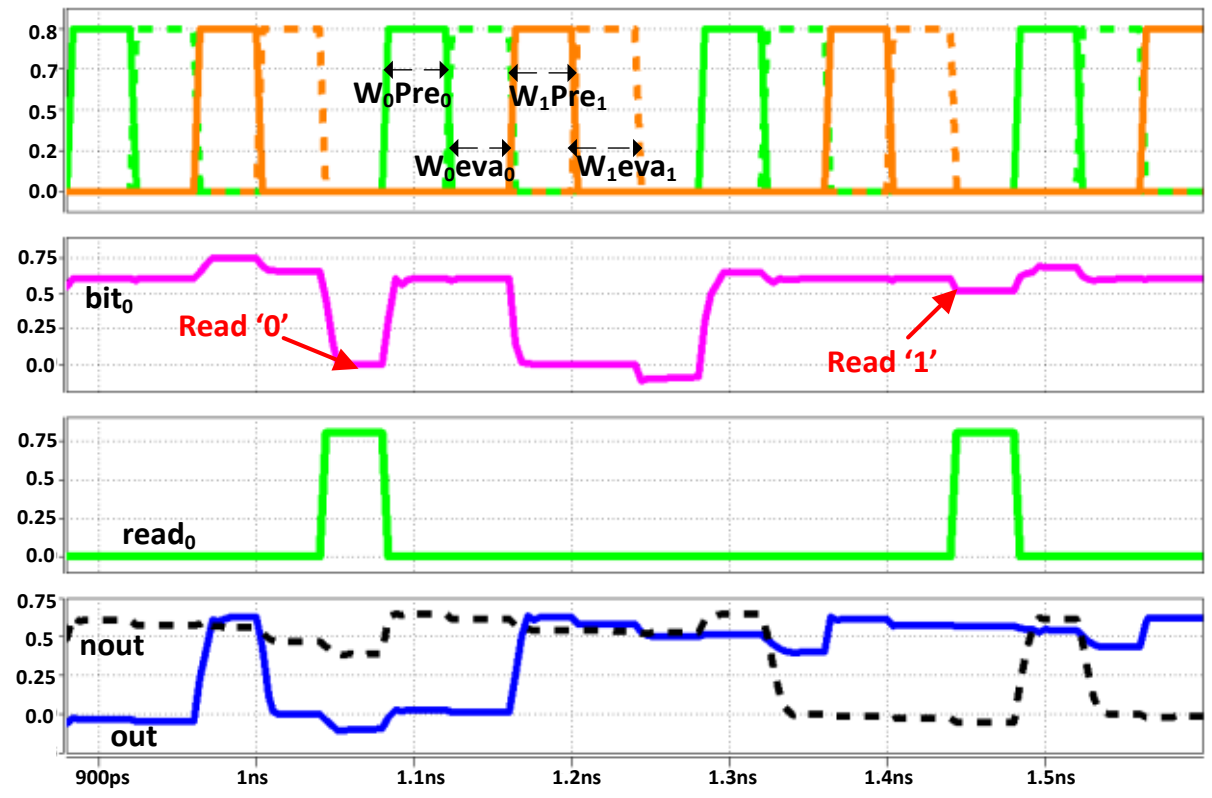

Fig. 5. Simulated waveforms of 10T-NWRAM operations

10T-NWRAM exploits the memory cache usage pattern that at a certain time activity is centered only on a small portion of memory block. During memory cell inactivity, all control signals are kept at ‘0’ allowing the cell to be in state-preserving mode; stored values are retained in parasitic capacitances of interconnect and adjacent transistors. However, due to leakage in xnwFETs, the stored charge could leak away over time. To restore the charges back to stored values, the control signals ($W_0pre_0$, $W_0pre_1$, $W_0eva_0$, $W_0eva_1$) are turned *ON* sequentially after a period of time. This restoring operation is shown in fig. 6 through HSPICE simulation results.

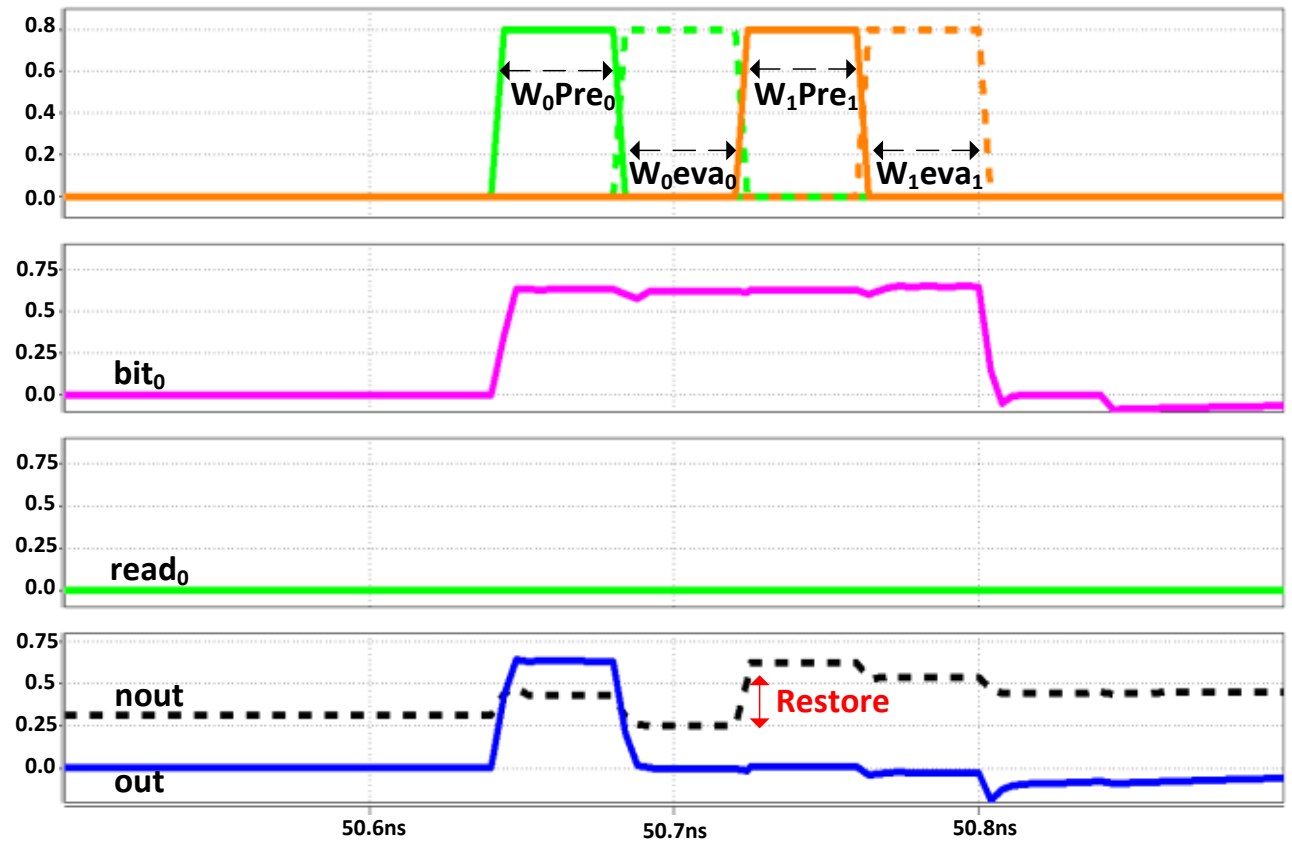

Fig. 6. Simulated waveforms of 10T-NWRAM operations

The physical layout of 10T-NWRAM is shown in Fig. 7. 10T-NWRAM follows a very regular grid based layout, which conforms to the N[3]ASIC fabric. Intra-cell routing is limited to only two layers of metal interconnect (M1 and M2).

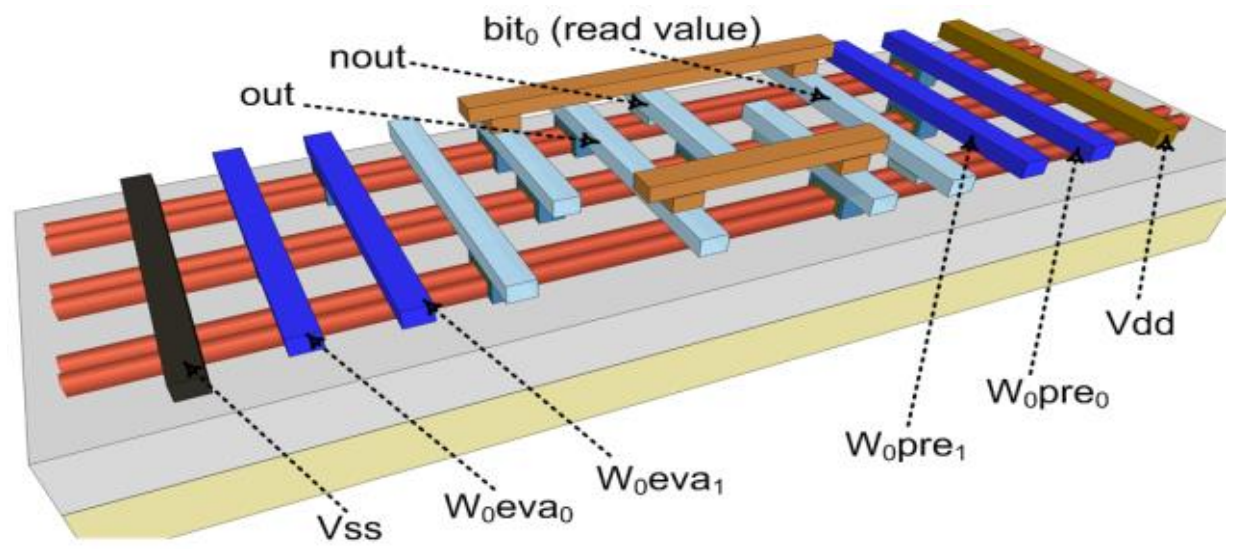

Fig. 7. 3-D layout of 10T-NWRAM memory cell

## IV. Benchmarking

In order to quantify the benefits of our 10T-NWRAM design over state-of-the-art SRAM designs, we have done extensive benchmarking. Layout analysis and HSPICE simulations were carried out to compare 10T NWRAM against high performance (HP) 6T-SRAM, low power (LP) 6T-SRAM, high performance gridded 8T-SRAM, and low power gridded 8T-SRAM in terms of cell area, active power, leakage power, read time and write time.

### A. Benchmarking Methodology

Fig. 8 show the layout of different memory cells used in this work. As shown in Fig. 8A, 10T-NWRAM follows a regular grid based layout with semiconducting nanowires, uniformly sized transistors and metal interconnects; therefore, 16nm 1D gridded design rules (Table I) from [1] [6] were used to calculate 10T-NWRAM cell area and interconnect parasitics.

TABLE I
1D Gridded Design Rules

| Pitch (16nm Tech) | **M1,M2 interconnect** | **Contact** |
|---|---|---|
| **1D-Gridded Design** | 60 ~ 40 nm | 50 nm |

2C-xnwFET devices, used for our 10T-NWRAM design were modeled and simulated using the TCAD Synopsys Sentaurus 3-D device simulator [2]. To model career transport in these devices, hydrodynamic charge transport model with quantum confinement correction was used [2]. Major xnwFET device characteristics are highlighted in table II. Table II also shows comparison of key device metrics with respect to PTM HP and LP device models during nominal conditions.

TABLE II
Comparison of Device Models

| | **2C-xnwFET** | **PTM_HP (NMOS)** | **PTM_LP (NMOS)** |
|---|---|---|---|
| **Ion** | $4.08^{-05}$ | $3.68^{-05}$ | $1.47^{-05}$ |
| **Ioff** | $1.56^{-09}$ | $1.05^{-08}$ | $1.99^{-12}$ |
| **Vdd (nominal)** | 0.8 | 0.7 | 0.9 |
| **Vth** | 0.27 | 0.47 | 0.68 |
| **Length/Width (nominal)** | 16/16 | 16/32 | 16/32 |

In order to do HSPICE circuit simulation of 10T-NWRAM, an HSPICE compatible device model was generated from TCAD simulations using the methodology described in [1]. Cell interconnect length and width were derived from cell layout (Fig. 1A), and PTM interconnect model [8] was used for interconnect RC calculations.

To scale the 6T-SRAM to 16nm technology node, we have collected published data about cell area and design rules from industry [9]-[1] for both high performance and low power 6T-

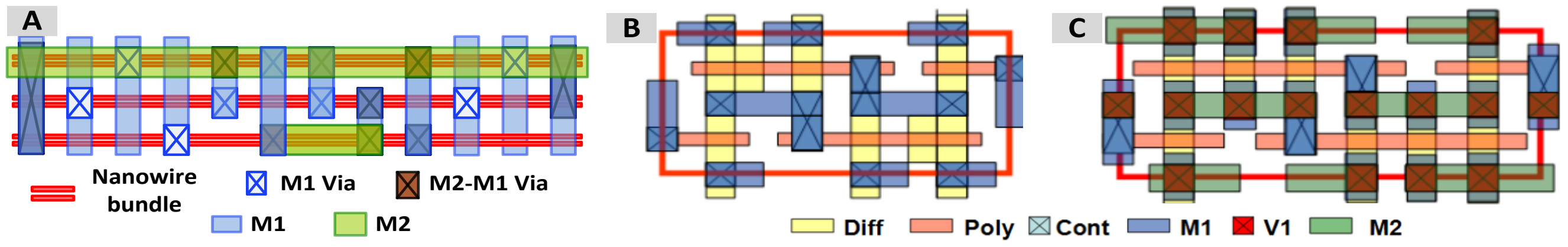

Fig. 8. Memory cell layout (2-D); A) 10T-NWRAM, B) 6T-SRAM [5], C) Gridded 8T-SRAM [5]

SRAM designs at 65nm, 45nm and 32nm technology nodes. From this data, various scaling factors were derived based on cell area, Poly, Metal1, Metal2 and Via scaling trends. These were used to calculate 6T-SRAM cell areas and corresponding 16nm design rules as shown in Table III and Table IV.

TABLE III
16 NM SCALED 6T-SRAM AREA

| Scaling factors | **2.45** | **2.02** | **1.75** | **1.64** |
|---|---|---|---|---|
| **Area in μm$^2$** | 0.028 | 0.042 | 0.056 | 0.064 |
| | 0.026 | 0.038 | 0.051 | 0.058 |
| | 0.025 | 0.037 | 0.049 | 0.056 |

TABLE IV
DESIGN RULES FOR 6T-SRAM

| Scaling factors | **1.31** | **1.42** | **1.38** | **1.31** |
|---|---|---|---|---|
| **M1x half pitch (nm)** | 32.49 | 27.5625 | 29.16 | 32.49 |
| | 28.88 | 24.5 | 25.92 | 28.88 |
| | 28.88 | 24.5 | 25.92 | 28.88 |
| **N+/P+ spacing (nm)** | 43.32 | 36.75 | 38.88 | 43.32 |
| | 33.7896 | 28.665 | 30.3264 | 33.7896 |
| | 37.544 | 31.85 | 33.696 | 37.544 |
| **Via spacing (nm)** | 32.49 | 27.5625 | 29.16 | 32.49 |
| | 28.88 | 24.5 | 25.92 | 28.88 |
| | 28.88 | 24.5 | 25.92 | 28.88 |

Interconnect lengths in 6T-SRAM cells were calculated from the layout (Fig. 8B) using cell area (Table III) and corresponding design rules (Table IV). Pass transistors and pull down transistors were considered to be 1.4x and 1.7x larger compared to pull-up transistors; PTM 16nm high performance and low power devices, and interconnect models [8] were used to simulate 6T-SRAM cell characteristics in HSPICE.

Similar simulations were carried out for manufacturing friendly gridded 8T-SRAM cell (Fig. 8C). 1-D gridded design rules shown in Table I, and 16nm PTM device and interconnect models [8] were used for simulations.

*B. Benchmarking Results*

Results from scaled memory cell area calculations are shown in Fig. 9. In this figure upper bound (colored black) and lower bound (colored yellow) corresponds to upper and lower bounds in cell area due to considered range (Table I) of design rules and scaling factors (Table III).

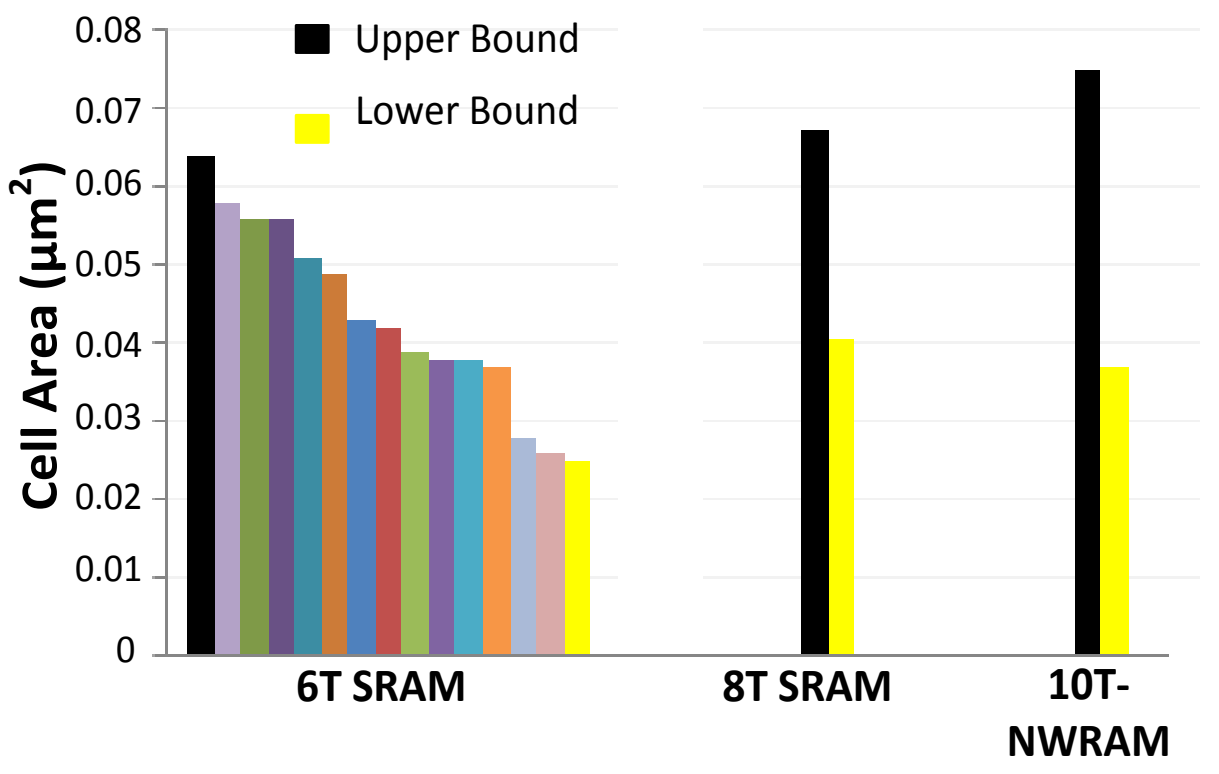

Fig. 9. Cell area comparison

Fig. 9 shows the lower bound of 10T-NWRAM cell area is comparable and in some cases better than scaled 6T-SRAM cells; whereas, the area comparisons between 8T-SRAM and 10T-NWRAM cells show similar results.

The upper bound in 10T-NWRAM shows larger area estimation for a single cell in comparison to upper bounds of SRAMs. This is mainly due to pessimistic assumptions in design rules during area calculations. The design rules for customized 10T-NWRAM cell is expected to be close to that of lower bound in table I, since 10T-NWRAM uses regular layout, uniformly sized transistors, and only two metal layers of interconnects.

10T-NWRAM write operation is significantly faster compared to SRAMs. Fig. 10 shows 10T-NWRAM write time to be almost 2x faster in comparison to HP 6T-SRAM and HP 8T-SRAM, and more than 4.5x faster when compared to LP 6T-SRAM and LP 8T-SRAM. This is primarily due to fast dynamic NAND logic style, high performance 2C-xnwFETs, and less load capacitance in the storage node. The load capacitance during bit transition is less in 10T-NWRAM because the bit transition in true and complementary nodes take place during non-overlapping clock cycles, whereas in SRAM both true and complementary bit values are flipped simultaneously.

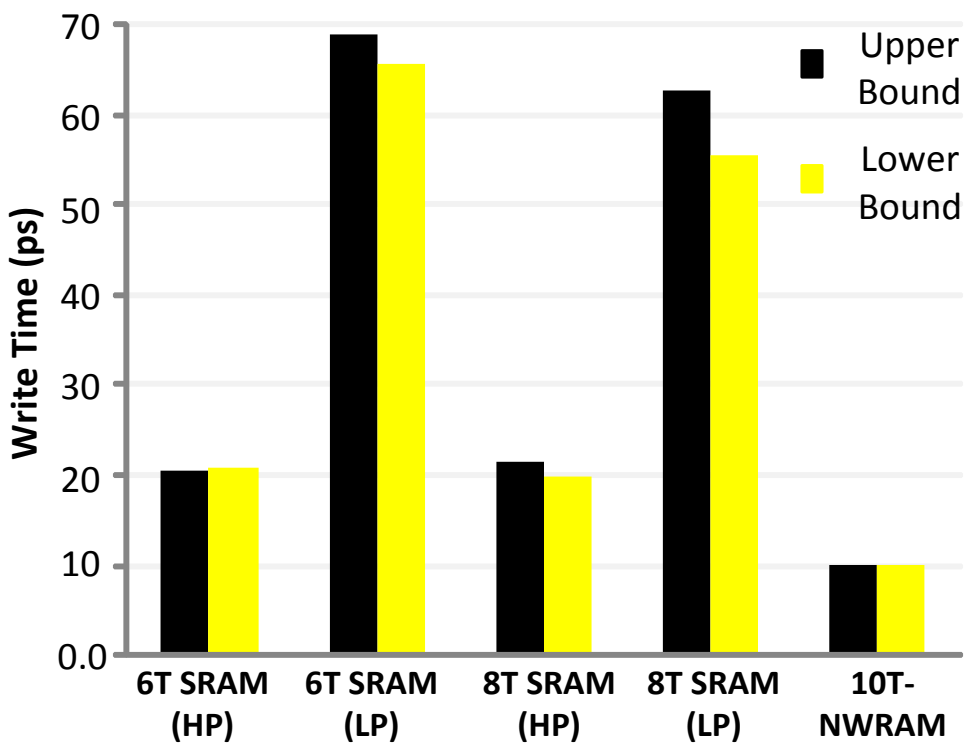

Fig. 10. Write time comparison

Fig. 11 shows the benchmarking results for read time. The read operation in 10T-NWRAM is faster (~2.7x over LP 6T-

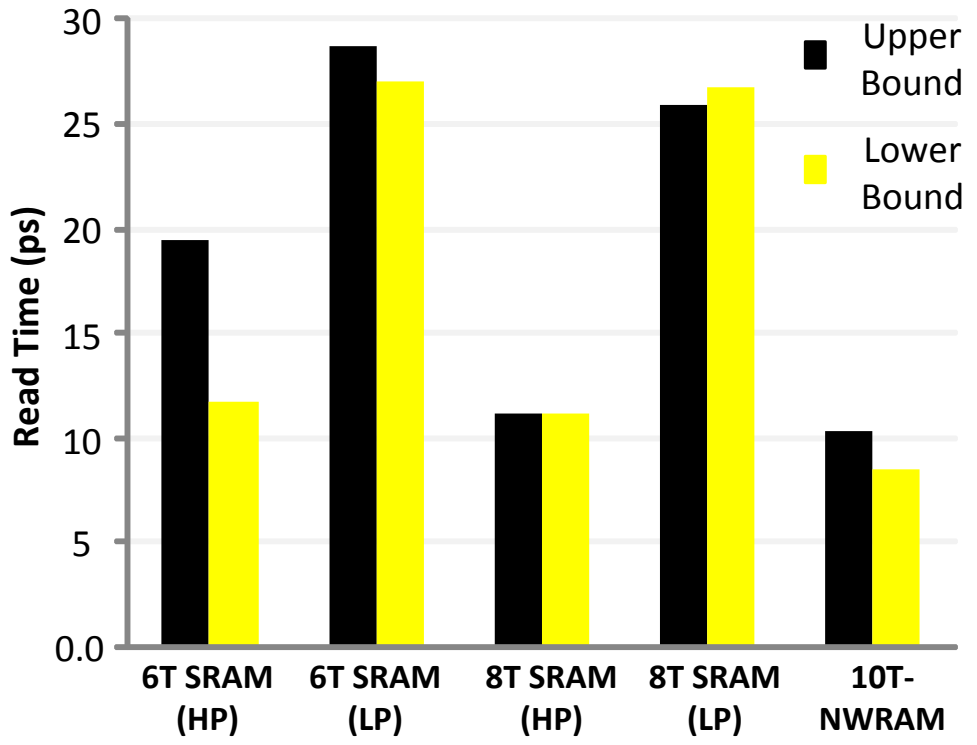

Fig. 11. Read time comparison

SRAM in the best case) compared to SRAMs due to the read logic scheme with data gating mechanism.

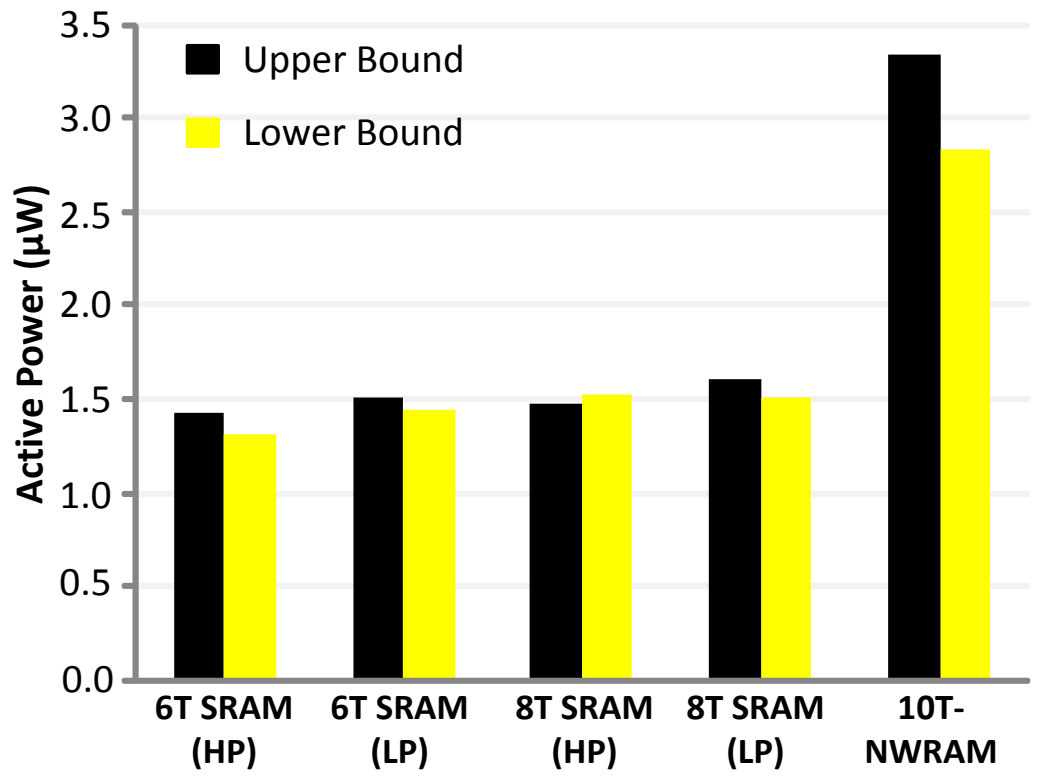

Fig. 12. Active power comparison

Power consumption results are shown in fig. 12 and fig. 13. 10T-NWRAM's active power per cell during read operation is higher (~2x) in comparison to SRAMs, which is primarily due to longer bitline length in 10T-NWRAM cell. During read operation, the $bit_0$ either remains charged or gets discharged depending on the stored value; the calculated read power consumption is the power consumed during $bit_0$ discharge. The physical layout of 10T-NWRAM is elongated in one direction: as shown in fig-8A, within the cell $bit_0$ propagation is horizontal through Metal2; whereas in SRAMs (Fig. 8B, 8C) $bit_0$ or $\sim bit_0$ propagation is vertical through Metal1, and the lengths are shorter, almost half compared to 10T-NWRAM $bit_0$ length. Therefore, the parasitic affect due to longer $bit_0$ length results in higher active power consumption for 10T-NWRAM cell.

Through memory array organization (i.e. more words, fewer bits in a block) the active power consumption of 10T-NWRAM block can be made similar to that of an SRAM block. In addition, layout optimizations to reduce $bit_0$ length and device optimizations to improve subthreshold characteristics may reduce active power consumption further.

The leakage power for 10T-NWRAM cell is significantly less compared to high performance SRAM designs. The lower bound data in fig. 13 shows 10T-NWRAM to be 35x and 14x better in terms of leakage when compared to HP 6T-SRAM and 8T-SRAM designs. This is mainly because of the state-preservation mechanism during idle period (as explained in section III) and transistor stacking effect in dynamic NAND gates.

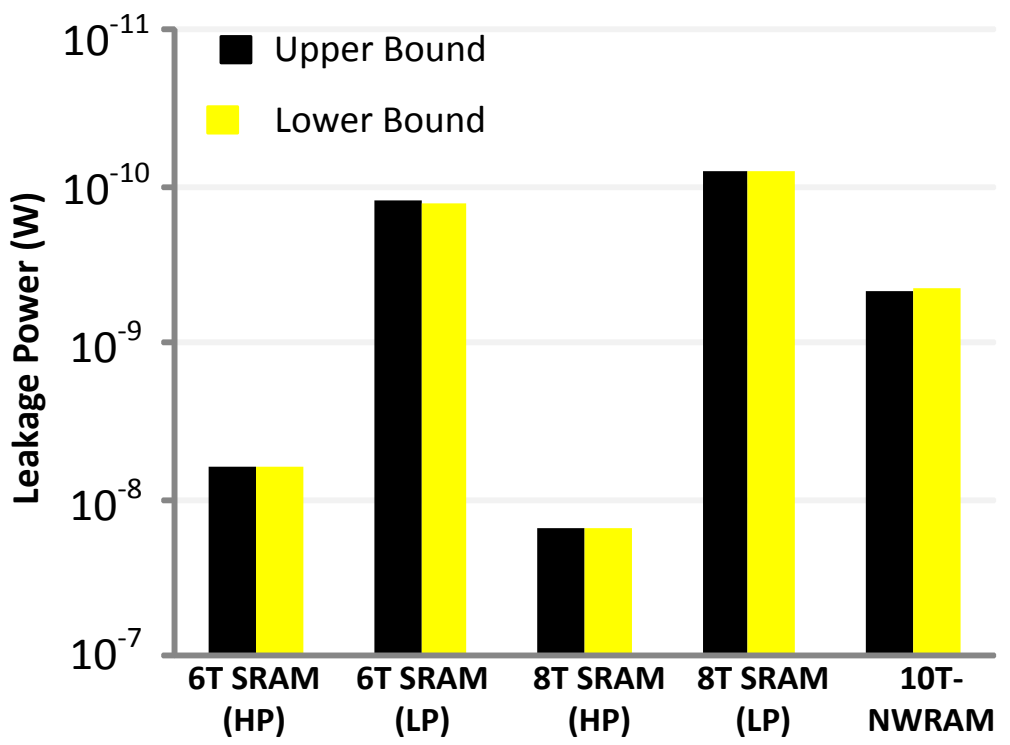

Fig. 13. Leakage power comparison

The leakage power of 10T-NWRAM cell can be further improved by optimizing the circuit to charge intermediate nodes. A different placement of control transistors (*pre*, *eva*) can allow charge sharing in transistor diffusion capacitances, which will increase the threshold voltage of transistors and pave the way for higher retention time in storage nodes.

The benefits of high density, high performance, and low leakage 10T-NWRAM, are also accomplished at a lower manufacturing complexity. In our previous work [1], we have shown that $N^3$ASIC based designs would have significantly less overlay requirements ($3\sigma = \pm 8$nm) for maximum yield in comparison to CMOS based designs ($3\sigma = \pm 3.3$nm).

## V. Conclusions

In this paper, details of 10T-NWRAM operation, physical layout and benchmarking were presented. In 10T-NWRAM, manufacturing complexities were significantly reduced due to the regular grid based cell layout, uniform devices and fewer metal interconnect layers. The benchmarking results show 10T-NWRAM is 2x faster and consumes 35x less leakage power when compared to gridded HP 8T-SRAM. The performance and leakage power benefits are also significant in comparison to HP and LP 6T-SRAM cell designs.